\documentstyle[12pt]{article}
\input{psfig}
\newlength{\dinwidth}
\newlength{\dinmargin}
\newcommand{\cm}      {\:\rm cm}
\newcommand{\GeV}      {\:\rm GeV}
\newcommand{\MeV}      {\:\rm MeV}
\begin{document}

\title{
{\bf Inclusive transverse momentum distributions
of charged particles in diffractive and non--diffractive
photoproduction at HERA}
       \\
\author
{\rm ZEUS Collaboration\\}
}
\date{ }
\maketitle
\vspace{5cm}

\begin{abstract}
Inclusive transverse momentum spectra of charged particles in
photoproduction events in the laboratory pseudorapidity range
$-1.2<\eta<1.4$ have been measured up to $p_{T}=8\GeV $ using the ZEUS
detector. Diffractive and non--diffractive reactions have been selected
with an average $\gamma p$ centre of mass (c.m.) energy of $\langle W
\rangle = 180\GeV$. For diffractive reactions, the $p_{T}$ spectra of the
photon dissociation events have been measured in two intervals of the
dissociated photon mass with mean values $\langle M_{X} \rangle = 5$ GeV
and $10$ GeV.  The inclusive transverse momentum spectra fall
exponentially in the low $p_{T}$ region. The non--diffractive data show a
pronounced high $p_{T}$ tail departing from the exponential shape.  The
$p_{T}$ distributions are compared to lower energy photoproduction data
and to hadron--hadron collisions at a similar c.m. energy.  The data are
also compared to the results of a next--to--leading order QCD calculation.
\end{abstract}

\vspace{-20cm}
\begin{flushleft}
\tt DESY 95-050 \\
March 1995 \\
\end{flushleft}

\setcounter{page}{0}
\thispagestyle{empty}

\newpage

%
\def\3{\ss}
\parindent 0cm
\footnotesize
\renewcommand{\thepage}{\Roman{page}}
\begin{center}
\begin{large}
The ZEUS Collaboration
\end{large}
\end{center}
M.~Derrick, D.~Krakauer, S.~Magill, D.~Mikunas, B.~Musgrave,
J.~Repond, R.~Stanek, R.L.~Talaga, H.~Zhang \\
{\it Argonne National Laboratory, Argonne, IL, USA}~$^{p}$\\[6pt]
R.~Ayad$^1$, G.~Bari, M.~Basile,
L.~Bellagamba, D.~Boscherini, A.~Bruni, G.~Bruni, P.~Bruni, G.~Cara
Romeo, G.~Castellini$^{2}$, M.~Chiarini,
L.~Cifarelli$^{3}$, F.~Cindolo, A.~Contin, M.~Corradi,
I.~Gialas$^{4}$,
P.~Giusti, G.~Iacobucci, G.~Laurenti, G.~Levi, A.~Margotti,
T.~Massam, R.~Nania, C.~Nemoz, \\
F.~Palmonari, A.~Polini, G.~Sartorelli, R.~Timellini, Y.~Zamora
Garcia$^{1}$,
A.~Zichichi \\
{\it University and INFN Bologna, Bologna, Italy}~$^{f}$
% ...
\\[6pt]
\clearpage
\hspace*{1mm}
$^{ 1}$ supported by Worldlab, Lausanne, Switzerland \\
\hspace*{1mm}
$^{ 2}$ also at IROE Florence, Italy  \\
\hspace*{1mm}
$^{ 3}$ now at Univ. of Salerno and INFN Napoli, Italy  \\
\hspace*{1mm}
$^{ 4}$ supported by EU HCM contract ERB-CHRX-CT93-0376 \\
\hspace*{1mm}
% ...
$^{16}$ now with OPAL Collaboration, Faculty of Physics at Univ. of
        Freiburg \\
% ...

\begin{tabular}{lp{15cm}}
$^{a}$ &supported by the Natural Sciences and Engineering Research
         Council of Canada (NSERC) \\
$^{b}$ &supported by the FCAR of Qu\'ebec, Canada\\
$^{c}$ &supported by the German Federal Ministry for Research and
         Technology (BMFT)\\
$^{d}$ &supported by the MINERVA Gesellschaft f\"ur Forschung GmbH,
         and by the Israel Academy of Science \\
$^{e}$ &supported by the German Israeli Foundation, and
         by the Israel Academy of Science \\
$^{f}$ &supported by the Italian National Institute for Nuclear Physics
         (INFN) \\
$^{g}$ &supported by the Japanese Ministry of Education, Science and
         Culture (the Monbusho)
         and its grants for Scientific Research\\
$^{h}$ &supported by the Korean Ministry of Education and Korea Science
         and Engineering Foundation \\
$^{i}$ &supported by the Netherlands Foundation for Research on Matter
         (FOM)\\
$^{j}$ &supported by the Polish State Committee for Scientific Research
         (grant No. SPB/P3/202/93) and the Foundation for Polish-
         German Collaboration (proj. No. 506/92) \\
$^{k}$ &supported by the Polish State Committee for Scientific
         Research (grant No. PB 861/2/91 and No. 2 2372 9102,
         grant No. PB 2 2376 9102 and No. PB 2 0092 9101) \\
$^{l}$ &partially supported by the German Federal Ministry for
         Research and Technology (BMFT) \\
$^{m}$ &supported by the German Federal Ministry for Research and
         Technology (BMFT), the Volkswagen Foundation, and the Deutsche
         Forschungsgemeinschaft \\
$^{n}$ &supported by the Spanish Ministry of Education and Science
         through funds provided by CICYT \\
$^{o}$ &supported by the Particle Physics and Astronomy Research
        Council \\
$^{p}$ &supported by the US Department of Energy \\
$^{q}$ &supported by the US National Science Foundation
\end{tabular}

\newpage
\pagenumbering{arabic}
\setcounter{page}{1}
\normalsize

\section{Introduction}
HERA is a prodigious source of quasi--real photons from reactions
where the electron is scattered at very small angles.
This permits the study of photoproduction reactions at photon--proton
centre of mass (c.m.) energies an order of magnitude larger than
in previous fixed target experiments.

The majority of the $\gamma p$ collisions are due to
interactions of the proton with the hadronic structure of the
photon, a process that has been successfully described by the
vector meson dominance model (VDM)\cite{VDM}.
Here, the photon is pictured to fluctuate into a virtual
vector meson that subsequently collides with the proton.
Such collisions exhibit the phenomenological characteristics of
hadron--hadron interactions.  In particular they
can proceed via diffractive or non--diffractive channels.
The diffractive interactions are characterized by very small
four momentum transfers and no colour exchange between the colliding
particles leading to final states where the colliding particles
appear either intact or as more massive dissociated states.

However, it has been previously
demonstrated that photoproduction collisions at high transverse
momentum cannot be described solely
in terms of the fluctuation of the photon into a hadron--like state
\cite{omega,na14}.  The deviations come from
contributions of two additional processes called
direct and anomalous.  In the former process
the photon couples directly to the charged
partons inside the proton.  The anomalous component corresponds to
the process where the photon couples to a {\it q\={q}} pair without
forming a bound state.  The interactions of the photon via the
hadron--like state and the anomalous component are
referred to as resolved photoproduction, since
both of them can be described in terms of the partonic structure of
the photon \cite{Storrow}.

In this paper we present the measurement of the transverse momentum
spectra of charged particles produced in
photoproduction reactions at an average c.m. energy of
$\langle W \rangle = 180\GeV$ and in the laboratory pseudorapidity
range $-1.2<\eta<1.4$
\footnote{Pseudorapidity $\eta$ is calculated from the relation
$ \eta = - ln( tan( \theta / 2))$, where
$\theta$ is a polar angle calculated with respect to the proton beam
direction.
}.
This range approximately corresponds to
the c.m. pseudorapidity interval of $0.8<\eta_{c.m.}<3.4$, where the
direction is defined such that positive
$\eta_{c.m.}$ values correspond to the photon fragmentation region.
The transverse momentum distributions of charged particles are studied
for non--diffractive and diffractive reactions separately.
The $p_{T}$ spectrum from non--diffractive events
is compared to low energy photoproduction data and to hadron--hadron
collisions at a similar c.m. energy.  In the region of high transverse
momenta we compare the data to the predictions of a next--to--leading order
QCD calculation.

The diffractive reaction ($\gamma p \rightarrow X p$), where $X$ results from
the dissociation of the photon,
was previously measured by the E612 Fermilab experiment
at much lower c.m. energies, $11.8<W<16.6\GeV$  \cite{chapin}.
It was demonstrated that the properties of the diffractive excitation of the
photon resemble diffraction of hadrons in terms of
the distribution of the dissociated mass, the distribution
of the four-momentum transfer between the colliding objects
\cite{chapin} and the ratio of the diffractive cross section
to the total cross section \cite{zeus-sigmatot}.
The hadronization of diffractively dissociated
photons has not yet been systematically studied.
In this analysis we present the measurement of inclusive $p_{T}$ spectra
in two intervals of the dissociated photon mass with mean values of
$\langle M_{X} \rangle = 5\GeV$ and $\langle M_{X} \rangle = 10\GeV$.

\section{Experimental setup}
The analysis is based on data collected with the ZEUS detector in 1993,
corresponding to an integrated luminosity of $0.40\:{\rm pb}^{-1}$.
The HERA machine was operating at an electron energy of $26.7\GeV$
and a proton energy of $820\GeV$, with 84 colliding
bunches.  In addition 10 electron and 6 proton bunches were left unpaired
for background studies (pilot bunches).

A detailed description of the ZEUS detector may be found elsewhere
\cite{status93,zeus-description}.
Here, only a brief description of the detector components used for
this analysis is given.
Throughout this paper the standard ZEUS right--handed
coordinate system is used,
which has its origin at the nominal interaction point.
The positive Z--axis points in the direction of the proton beam,
called the forward direction, and X points towards the centre of the
HERA ring.

Charged particles created in $ep$ collisions are tracked
by the inner tracking detectors which operate in a
magnetic field of $1.43{\: \rm T}$ provided by a thin superconducting
solenoid.
Immediately surrounding the beampipe is the vertex detector (VXD),
a cylindrical drift chamber
which consists of 120 radial cells, each with 12 sense wires running
parallel to the beam axis \cite{VXD}.
The achieved resolution is
$50\:{\rm \mu m}$ in the central region of a cell and  $150\:{\rm \mu m}$
near the edges.  Surrounding the VXD is the central tracking detector
(CTD) which consists of 72 cylindrical drift chamber layers organized in 9
superlayers \cite{CTD}.  These superlayers alternate between those with
wires parallel to the collision axis and those with wires inclined at
a small angle to provide a stereo view.  The magnetic field is significantly
inhomogeneous towards the ends of the CTD thus complicating the electron drift.
With the present understanding of the chamber, a
spatial resolution of $\approx 260\:{\rm \mu m}$ has been achieved.
The hit efficiency of the chamber is greater than $95\%$.

In events with charged particle tracks, using the combined data from both
chambers, the position resolution of the
reconstructed primary vertex are $0.6\:{\rm cm}$
in the Z direction and $0.1\:{\rm cm}$ in the XY plane.
The resolution in transverse momentum for full length tracks is
 $\sigma_{p_{T}} / p_{T} \leq 0.005 \cdot p_{T} \oplus 0.016$ ($p_{T}$ in
$\GeV$).
The description of the track and the vertex reconstruction algorithms
may be found in \cite{zeus-breit} and references therein.

The solenoid is surrounded by the high resolution
uranium--scintillator calorimeter (CAL) divided into the forward (FCAL), barrel
(BCAL) and rear (RCAL) parts \cite{CAL}.
Holes of $20 \times 20 {\rm\: cm}^{2}$ in
the centre of FCAL and RCAL are required to accommodate the HERA beam
pipe.
Each of the calorimeter parts is subdivided into towers which in
turn are segmented longitudinally into electromagnetic (EMC) and
hadronic (HAC) sections.  These sections are further subdivided
into cells, which are read out by two photomultiplier tubes.
Under test beam conditions, an energy resolution of the calorimeter of
$\sigma_{E}/E = 0.18/\sqrt{E (\GeV)}$ for electrons and
$\sigma_{E}/E = 0.35/\sqrt{E (\GeV)}$ for hadrons was measured.
In the analysis presented here CAL cells with an EMC (HAC) energy below
$60\MeV$ ($110\MeV$) are excluded to minimize the effect of calorimeter
noise.  This noise is dominated by uranium activity and has an
r.m.s.~value below $19\MeV$ for EMC cells and below $30\MeV$ for HAC
cells.

The luminosity detector (LUMI) measures the rate of the
Bethe--Heitler process $e p \rightarrow e \gamma p$.
The detector consists of two lead--scintillator sandwich
calorimeters installed in the HERA tunnel and
is designed to detect electrons scattered at very small
angles and photons emitted along the electron beam direction \cite{lumi}.
Signals in the LUMI electron calorimeter are used to tag events and to measure
the energy of the interacting photon, $E_{\gamma}$, from
$E_{\gamma}=E_{e}-E'_{e}=26.7\GeV-E'_{e}$, where $E'_{e}$ is the
energy measured in the LUMI.

\section{Trigger}
The events used in the following analysis were collected using
a trigger requiring a  coincidence of the signals in the LUMI electron
calorimeter and in the central calorimeter.
The small angular acceptance of the LUMI electron calorimeter
implied that in all the triggered events
the virtuality of the exchanged photon was between
$4\cdot 10^{-8} < Q^{2}< 0.02 \GeV^{2}$.

The central calorimeter trigger required an energy deposit in the
RCAL EMC section of more than $464\:{\rm MeV}$
(excluding the towers immediately adjacent to the beam pipe)
or $1250\:{\rm MeV}$ (including those towers).
In addition we also used the
events triggered by an energy in the BCAL EMC section exceeding
$3400\:{\rm MeV}$.
At the trigger level the energy was calculated using only towers
with more than $464\:{\rm MeV}$ of deposited energy.

\section{Event selection}
In the offline analysis the energy
of the scattered electron detected in the LUMI calorimeter was required to
satisfy $15.2<E'_{e}<18.2\GeV$, limiting the $\gamma p$ c.m. energy
to the interval $167<W<194\GeV$.
The longitudinal vertex position determined from tracks
was required to be $-35\cm < Z_{vertex} < 25\cm $.
The vertex cut removed a substantial part of the
beam gas background and limited the data sample to
the region of  uniform detector acceptance.
The cosmic ray background was suppressed by requiring
the transverse momentum imbalance of the deposits in the main calorimeter,
$P_{missing}$,
relative to the square root of the total transverse energy, $\sqrt{E_{T}}$, to
be small: $P_{missing}/\sqrt{E_{T}} < 2\sqrt{\GeV}$.

The data sample was divided into
a diffractive and a non--diffractive
subset according to the pseudorapidity,
$\eta_{max}$, of the most forward
energy deposit in the FCAL with energy above $400\MeV$.
The requirement of $\eta_{max} < 2$ selects events with a pronounced
rapidity gap
that are predominantly due to diffractive processes
($\approx 96\%$ according to Monte Carlo (MC) simulation, see section~6).
The events with $\eta_{max} > 2$
are almost exclusively ($\approx 95\%$) due to non-diffractive
reactions.
The final non--diffractive data sample consisted of 149500 events.

For the diffractive data sample ($\eta_{max} < 2$)
an additional cut $\eta_{max} > -2$ was applied to suppress
the production of light vector mesons $V$ in the diffractive reactions
 $\gamma p \rightarrow V p$ and
 $\gamma p \rightarrow V N$, where $N$ denotes a nucleonic
system resulting from the dissociation of the proton.
The remaining sample was analyzed as a function of the mass of the
dissociated system reconstructed from the empirical relationship
\[ M_{X rec} \approx A\cdot\sqrt{E^{2}-P_{Z}^{2}}+B =
A\cdot\sqrt{(E+P_{Z}) \cdot E_{\gamma}}+B .\]
The above formula exploits the fact that in tagged photoproduction
the diffractively excited photon state has a relatively small
transverse momentum.
The total hadronic energy, $E$, and longitudinal
momentum $P_{Z}=E \cdot cos\theta$
were measured with the uranium calorimeter
by summing over all the energy deposits of at least $160\MeV$.
The correction factors $A=1.7$ and $B=1.0\GeV$
compensate for the effects of energy loss in
the inactive material, beam pipe holes,
and calorimeter cells that failed the energy threshold cuts.
The formula was optimized to give the best approximation of
the true invariant mass in diffractive photon dissociation events
obtained from MC simulations, while being insensitive to the
calorimeter noise.
The diffractive data were
analyzed in two intervals of the reconstructed mass,
namely \nobreak{$4<M_{X\:rec}<7\GeV$} and $8<M_{X\:rec}<13\GeV$.
According to the MC simulation
the first cut selects events generated with a mass having a mean
value and spread of $\langle M_{X\:GEN} \rangle = 5\GeV$ and
${\rm r.m.s.} = 1.8\GeV$. The second cut results in
$\langle M_{X GEN} \rangle = 10\GeV$ and ${\rm r.m.s.} = 2.3\GeV$.
Details of the MC simulation are given in section \ref{s:mc}.
The final data sample consisted of
5123 events in the lower $M_{X}$ interval and of 2870 events in the upper
interval.

The contamination of the final data samples from e--gas background ranges
from \nobreak{$<0.1\%$} (non-diffractive sample) to
 $\approx 10\%$ (diffractive sample, $\langle M_{X} \rangle = 5\GeV$).
The p--gas contribution is between $1\%$ (non-diffractive sample) and
 $2\%$ (diffractive sample, $\langle M_{X} \rangle = 5\GeV$).
The e--gas background was statistically subtracted using
the electron pilot bunches.
A similar method was used
to correct for the p--gas background that
survived the selection cuts because of an accidental coincidence with
an electron bremsstrahlung $(ep \rightarrow \gamma ep)$.
A large fraction of these background events were identified using
the LUMI detector, since the energy deposits in the electron and
photon calorimeters summed up to the electron beam energy.
The identified background events were
included with negative weights into all of the distributions
in order to compensate for the unidentified part of the
coincidence background.
A detailed description of the statistical background
subtraction method may be found in \cite{zeus-sigmatot,phdburow}.

\section{Track selection}

The charged tracks used for this analysis were selected with the
following criteria:\
\begin{itemize}
\item
only tracks accepted by an event vertex fit were selected.
This eliminated most of the tracks that
came from secondary interactions and decays of short lived particles;
\item
tracks must have hits in each of the first 5 superlayers of the CTD.
This requirement ensures that only long, well reconstructed
tracks are used for the analysis;
\item
$-1.2<\eta<1.4$ and $p_{T} > 0.3\GeV$.
These two cuts select the region of high
acceptance of the CTD where the detector response and systematics are best
understood.
\end{itemize}
Using Monte Carlo events, we estimated that
the efficiency of the charged track reconstruction
convoluted with the acceptance of the selection
cuts is about $90\%$ and is uniform in $p_{T}$.
The contamination of the final sample from
secondary interaction tracks,  products of decays of short lived particles,
and from spurious tracks (artifacts of the reconstruction algorithm)
ranges from $5\%$ at $p_{T}=0.3\GeV$ to $3\%$ for $p_{T}>1\GeV$.
The inefficiency and remaining contamination of the final track sample
is accounted for by the acceptance correction described in the
following section.

The transverse momenta of the measured tracks displayed no
correlation with $\eta$ over the considered interval and were
symmetric with respect to the charge assigned to the track.

\section{Monte Carlo models}
\label{s:mc}
For the acceptance correction and selection cut validation we used
Monte Carlo events generated with a variety of programs.
Soft, non-diffractive collisions of the proton with a
VDM type photon were generated using HERWIG 5.7 with the
minimum bias option \cite{HERWIG}.  The generator was tuned to fit the
ZEUS data on charged particle multiplicity and transverse energy
distributions.
For the evaluation of the model dependence of our measurements
we also used events from the PYTHIA generator with the
soft hadronic interaction option \cite{PYTHIA}.
Hard resolved and direct subprocesses were simulated using the standard
HERWIG 5.7 generator with the lower cut-off on the transverse momentum of the
final--state partons, $p_{T min}$, chosen to be $2.5 \GeV$.
For the parton densities of the colliding particles, the GRV--LO \cite{GRV}
(for the photon) and MRSD$'$\_ \cite{MRS} (for the proton)
parametrisations were used.
As a cross--check we also used hard $\gamma p$ scattering events
generated by
PYTHIA with $p_{T min} = 5\GeV$.
The soft and hard MC components were
combined in a ratio that gave the best description of
the transverse momentum distribution
of the track with the largest $p_{T}$ in each event.
For $p_{T min} = 2.5\GeV$,
the hard component comprises $11\%$ of the non-diffractive sample and
for $p_{T min} = 5\GeV$ only about $3\%$.

Each diffractive subprocess was generated separately.
The diffractive production of vector mesons $(\rho, \omega, \phi)$
was simulated with PYTHIA.
The same program was used to simulate the double diffractive
dissociation ($\gamma p \rightarrow X N$).  The
diffractive excitation of the photon ($\gamma p \rightarrow X p$)
was generated with the EPDIF
program which models the diffractive system as a
quark--antiquark pair produced along the collision axis \cite{Solano}.
Final state QCD radiation and hadronization were simulated using JETSET
\cite{PYTHIA}.
For the study of systematic uncertainties, a similar sample of events
was obtained by enriching the standard PYTHIA diffractive events
with the hard component simulated using the POMPYT Monte Carlo program
(hard, gluonic pomeron with the direct photon option)
\cite{bruni}.

The MC samples corresponding to the diffractive subprocesses were
combined with the non-diffractive component in the proportions
given by the ZEUS measurement of the partial photoproduction
cross sections \cite{zeus-sigmatot}.  The MC events were generated without
electroweak
radiative corrections.  In the considered $W$ range, the QED radiation effects
result in $\approx 2\%$ change in the number of measured events so that
the effect on the results of this analysis are negligible.
The generated events were processed through the detector and trigger
simulation programs and run through the standard ZEUS reconstruction
chain.

\section{Acceptance correction}
\label{s:correction}
The acceptance corrected transverse momentum spectrum was
 derived from the reconstructed spectrum of charged tracks,
by means of a
multiplicative correction factor, calculated using Monte Carlo techniques:
\[ C(p_{T}) =
(\frac{1}{N_{gen\: ev}} \cdot \frac{d N_{gen}}{d p_{T gen}})/
(\frac{1}{N_{rec\: ev}} \cdot \frac{d N_{rec}}{d p_{T rec}}) .
\]
$N_{gen}$ denotes the number of primary charged particles generated
with a transverse momentum $p_{T gen}$ in the considered
pseudorapidity interval and $N_{gen\: ev}$ is the number of generated
events.  Only the events corresponding to the appropriate
type of interaction were included, e.g. for the
lower invariant mass interval of the diffractive sample
only the Monte Carlo events corresponding to diffractive photon dissociation
with the generated invariant mass $4<M_{X gen}<7\GeV$ were used.
$N_{rec}$ is the number of reconstructed tracks passing
the experimental cuts with a reconstructed transverse momentum
of $p_{T rec}$, while $N_{rec\: ev}$ denotes the number of events used.
Only the events passing the trigger simulation and the
experimental event selection criteria were included in the calculation.
To account for the contribution
of all the subprocesses, the combination of the MC samples described
in section~\ref{s:mc} was used.

This method corrects for the following effects in the data:
\begin{itemize}
\item
the limited trigger acceptance;
\item
the inefficiencies of the event selection cuts, in particular
the contamination of the diffractive spectra from
non--diffractive processes and the events with
a dissociated mass that was incorrectly reconstructed.
Also the non-diffractive sample is corrected for the
contamination from diffractive events with high dissociated mass;
\item
limited track finding efficiency and acceptance of the
track selection cuts, as well as the limited resolution in
momentum and angle;
\item
loss of tracks due to secondary
interactions and contamination from secondary tracks;
\item
decays of charged pions and kaons, photon conversions and decays of
lambdas and neutral kaons.  Thus, in the final spectra the
charged kaons appear, while the decay products
of neutral kaons and lambdas do not.  For all
the other strange and charmed states, the decay products were included.
\end{itemize}

The validity of our acceptance correction method relies on the
correct simulation of the described effects in the Monte Carlo program.
The possible discrepancies between reality and Monte Carlo simulation were
analyzed and the estimation of the effect on the final
distributions was included in the systematic uncertainty, as
described in the following section.

\section{Systematic effects}
\label{s:systematics}
One of the potential sources of systematic inaccuracy
is the tracking system and its
simulation in the Monte Carlo events used for the acceptance correction.  Using
an alternative simulation code with artificially degraded tracking
performance we verified that the efficiency to find a
track which fulfills all the selection cuts is known with an accuracy of
about $10\%$.
The error due to an imprecise description of the momentum resolution at
high $p_{T}$ is negligible compared to the statistical precision of the data.
We also verified that the final spectra would not change significantly
if the tracking resolution at high $p_{T}$ had non--gaussian tails
at the level of $10\%$
or if the measured momentum was systematically shifted from the true
value by the momentum resolution.

Another source of systematic uncertainty is the Monte Carlo
simulation of the trigger response.
We verified that even a very large ($20\%$)
inaccuracy of the BCAL energy threshold
would not produce a statistically significant effect.  An incorrect
RCAL trigger simulation would change the number of events
observed, but would not affect the final $p_{T}$ spectra since it
is normalized to the number of events.  The correlation between
the RCAL energy and the $p_{T}$ of tracks is very small.

To evaluate the model dependence we repeated the
calculation of the correction factors
using an alternative set of Monte Carlo programs (see section
\ref{s:mc}) and compared the results with the original ones.
The differences between the obtained factors
varied between $5\%$ for the high mass diffractive sample
and $11\%$ for the non--diffractive one.
The sensitivity of the result to the assumed relative
cross sections of the physics processes was checked by varying the
subprocess ratios within the error limits given in \cite{zeus-sigmatot}.
The effect was at most $3\%$.

All the above effects were combined in quadrature, resulting in an
overall systematic uncertainty of the charged particle rates
as follows: $15\%$ in the non-diffractive sample,
$15\%$ in the $\langle M_{X} \rangle = 5\GeV$ diffractive sample and
$9\%$ in the $\langle M_{X} \rangle = 10\GeV$ diffractive sample.
All these systematic errors are independent of $p_{T}$.

\section{Results}
\begin{figure}[h]
\centerline{\hbox{
\psfig{figure=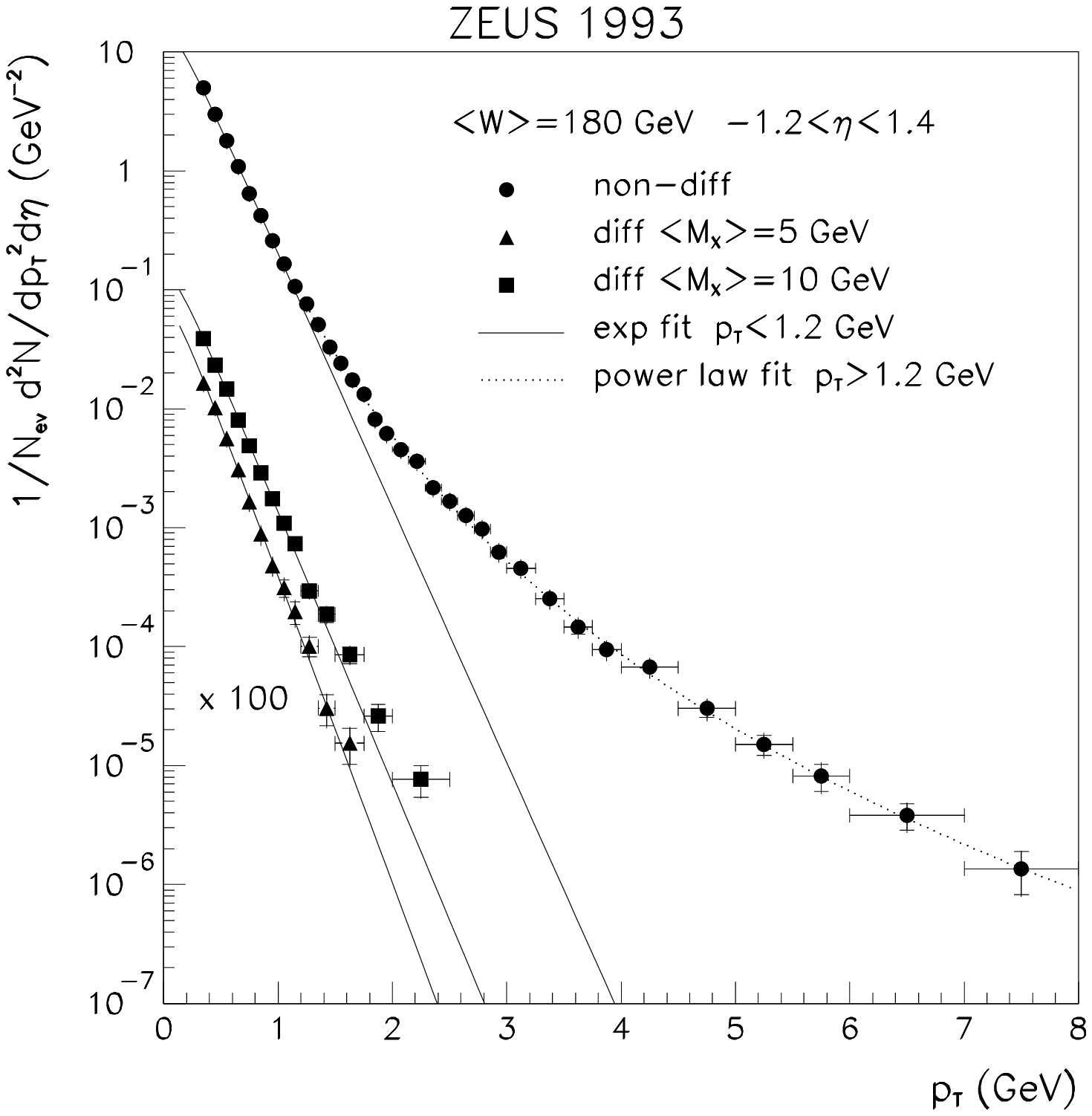}}}
\bf\caption{\it
Inclusive transverse momentum distributions of charged particles in
photoproduction events at $\langle W \rangle = 180\GeV$
averaged over the pseudorapidity interval of $-1.2<\eta<1.4$.
The inner error bars indicate the
statistical errors and the outer ones represent
 the quadratic sum of the statistical and
systematic errors.  Solid lines indicate fits of equation
(\protect\ref{e:exp}) to the
data in the region of $p_{T} < 1.2\GeV$.  The dotted line shows a
power law formulae (\protect\ref{e:power})
fitted to the non--diffractive data for $p_{T} > 1.2\GeV$.
For the sake of clarity the diffractive points are shifted
down by two orders of magnitude.
}
\label{f:corrected_pt}
\end{figure}

The double differential rate of charged particle
production in an event of a given type is calculated as the
number of charged particles $\Delta N$ produced
within $\Delta \eta$ and $\Delta p_{T}$ in $N_{ev}$
events as a function of $p_{T}$:
\[\frac{1}{N_{ev}} \cdot \frac{d^{2}N}{dp^{2}_{T} d\eta} =
  \frac{1}{N_{ev}} \cdot
  \frac{1}{2 p_{T} \Delta \eta} \cdot
  \frac{\Delta N}{\Delta p_{T}} .\]
The charged particle transverse momentum spectrum was derived from the
transverse momentum distribution of observed tracks normalized
to the number of data events by means of the
correction factor described in section \ref{s:correction}.
The resulting charged particle production rates in
diffractive and non-diffractive events are presented in
Fig.~\ref{f:corrected_pt} and
listed in Tables \ref{t:results1}, \ref{t:results2} and \ref{t:results3}.
In the figure the inner error bars indicate the
statistical error.  Quadratically combined statistical and
systematic uncertainties are shown as the outer error bars.
The $\langle M_{X} \rangle = 5\GeV$ diffractive spectrum extends
to $p_{T}=1.75\GeV$ and the $\langle M_{X} \rangle = 10\GeV$ distribution
extends to $p_{T}=2.5\GeV$.
The non--diffractive distribution falls steeply
in the low $p_{T}$ region but lies above the exponential fit
at higher $p_{T}$ values.  The measurements extend to $p_{T}=8\GeV$.

\begin{table}[h]
\centerline{\hbox{
\begin{tabular}{|c||c|c|c|}
\hline
\ \ $p_{T} [\GeV]$ \ \ &
$\frac{1}{N_{ev}} \cdot \frac{d^{2}N}{dp^{2}_{T} d\eta} [\GeV^{-2}]$ &
$\sigma_{stat} [\GeV^{-2}]$ &
$\sigma_{syst} [\GeV^{-2}]$\\
\hline\hline
  0.30-- 0.40 &  4.98 &  0.05 &  0.74 \\
  0.40-- 0.50 &  2.99 &  0.03 &  0.44 \\
  0.50-- 0.60 &  1.78 &  0.02 &  0.26 \\
  0.60-- 0.70 &  1.09 &  0.01 &  0.16 \\
  0.70-- 0.80 &  0.641 &  0.012 &  0.096 \\
  0.80-- 0.90 &  0.420 &  0.010 &  0.063 \\
  0.90-- 1.00 &  0.259 &  0.007 &  0.038 \\
  1.00-- 1.10 &  0.164 &  0.005 &  0.024 \\
  1.10-- 1.20 &  0.107 &  0.004 &  0.016 \\
  1.20-- 1.30 &  0.0764 &  0.0034 &  0.0114 \\
  1.30-- 1.40 &  0.0513 &  0.0017 &  0.0077 \\
  1.40-- 1.50 &  0.0329 &  0.0012 &  0.0049 \\
  1.50-- 1.60 &  0.0242 &  0.0010 &  0.0036 \\
  1.60-- 1.70 &  0.0175 &  0.0008 &  0.0026 \\
  1.70-- 1.80 &  0.0133 &  0.0006 &  0.0020 \\
  1.80-- 1.90 &  0.0082 &  0.0005 &  0.0012 \\
  1.90-- 2.00 &  0.00615 &  0.00038 &  0.00092 \\
  2.00-- 2.14 &  0.00454 &  0.00028 &  0.00068 \\
  2.14-- 2.29 &  0.00360 &  0.00024 &  0.00054 \\
  2.29-- 2.43 &  0.00215 &  0.00017 &  0.00032 \\
  2.43-- 2.57 &  0.00166 &  0.00013 &  0.00025 \\
  2.57-- 2.71 &  0.00126 &  0.00012 &  0.00018 \\
  2.71-- 2.86 &  0.00098 &  0.00010 &  0.00015 \\
  2.86-- 3.00 &  0.000625 &  0.000071 &  0.000093 \\
  3.00-- 3.25 &  0.000456 &  0.000048 &  0.000068 \\
  3.25-- 3.50 &  0.000252 &  0.000031 &  0.000037 \\
  3.50-- 3.75 &  0.000147 &  0.000020 &  0.000022 \\
  3.75-- 4.00 &  0.000094 &  0.000012 &  0.000014 \\
  4.00-- 4.50 &  0.000067 &  0.000008 &  0.000010 \\
  4.50-- 5.00 &  0.0000301 &  0.0000045 &  0.0000045 \\
  5.00-- 5.50 &  0.0000151 &  0.0000029 &  0.0000023 \\
  5.50-- 6.00 &  0.0000082 &  0.0000021 &  0.0000012 \\
  6.00-- 7.00 &  0.0000038 &  0.0000009 &  0.0000006 \\
  7.00-- 8.00 &  0.0000014 &  0.0000005 &  0.0000002 \\
\hline
\end{tabular}
}}
\vspace{1cm}
\bf\caption{\it
The rate of charged particle production in an average
non--diffractive event.  The data correspond to $-1.2<\eta<1.4$.
The $\sigma_{stat}$ and $\sigma_{syst}$ denote the statistical and
systematic errors.}
\label{t:results1}
\end{table}

\begin{table}[h]
\centerline{\hbox{
\begin{tabular}{|c||c|c|c|}
\hline
\ \ $p_{T} [\GeV]$ \ \ &
$\frac{1}{N_{ev}} \cdot \frac{d^{2}N}{dp^{2}_{T} d\eta} [\GeV^{-2}]$ &
$\sigma_{stat} [\GeV^{-2}]$ &
$\sigma_{syst} [\GeV^{-2}]$\\
\hline\hline
  0.30-- 0.40 &  1.63 &  0.06 &  0.24 \\
  0.40-- 0.50 &  1.02 &  0.04 &  0.15 \\
  0.50-- 0.60 &  0.559 &  0.028 &  0.083 \\
  0.60-- 0.70 &  0.308 &  0.019 &  0.046 \\
  0.70-- 0.80 &  0.165 &  0.013 &  0.024 \\
  0.80-- 0.90 &  0.088 &  0.011 &  0.013 \\
  0.90-- 1.00 &  0.0479 &  0.0059 &  0.0071 \\
  1.00-- 1.10 &  0.0312 &  0.0052 &  0.0046 \\
  1.10-- 1.20 &  0.0196 &  0.0042 &  0.0029 \\
  1.20-- 1.35 &  0.0100 &  0.0018 &  0.0015 \\
  1.35-- 1.50 &  0.00304 &  0.00087 &  0.00045 \\
  1.50-- 1.75 &  0.00153 &  0.00052 &  0.00023 \\
\hline
\end{tabular}
}}
\vspace{1cm}
\bf\caption{\it
The rate of charged particle production in an average
event with a diffractively dissociated photon state
of a mass $\langle M_{X} \rangle = 5\GeV$.
The data correspond to $-1.2<\eta<1.4$.
The $\sigma_{stat}$ and $\sigma_{syst}$ denote the statistical and
systematic errors.}
\label{t:results2}
\end{table}

\begin{table}[h]
\centerline{\hbox{
\begin{tabular}{|c||c|c|c|}
\hline
\ \ $p_{T} [\GeV]$ \ \ &
$\frac{1}{N_{ev}} \cdot \frac{d^{2}N}{dp^{2}_{T} d\eta} [\GeV^{-2}]$ &
$\sigma_{stat} [\GeV^{-2}]$ &
$\sigma_{syst} [\GeV^{-2}]$\\
\hline\hline
  0.30-- 0.40 &  3.87 &  0.10 &  0.34 \\
  0.40-- 0.50 &  2.32 &  0.06 &  0.20 \\
  0.50-- 0.60 &  1.46 &  0.04 &  0.13 \\
  0.60-- 0.70 &  0.803 &  0.033 &  0.072 \\
  0.70-- 0.80 &  0.485 &  0.023 &  0.043 \\
  0.80-- 0.90 &  0.288 &  0.017 &  0.025 \\
  0.90-- 1.00 &  0.176 &  0.012 &  0.015 \\
  1.00-- 1.10 &  0.109 &  0.009 &  0.009 \\
  1.10-- 1.20 &  0.0732 &  0.0075 &  0.0065 \\
  1.20-- 1.35 &  0.0294 &  0.0035 &  0.0026 \\
  1.35-- 1.50 &  0.0186 &  0.0028 &  0.0016 \\
  1.50-- 1.75 &  0.0086 &  0.0014 &  0.0008 \\
  1.75-- 2.00 &  0.00260 &  0.00066 &  0.00023 \\
  2.00-- 2.50 &  0.00076 &  0.00023 &  0.00007 \\
\hline
\end{tabular}
}}
\vspace{1cm}
\bf\caption{\it
The rate of charged particle production in an average
event with a diffractively dissociated photon state
of a mass $\langle M_{X} \rangle = 10\GeV$.
The data correspond to $-1.2<\eta<1.4$.
The $\sigma_{stat}$ and $\sigma_{syst}$ denote the statistical and
systematic errors.}
\label{t:results3}
\end{table}

The soft interactions of hadrons can be successfully described by
thermodynamic models that predict a steep fall of the transverse
momentum spectra that can be approximated with the exponential form
\cite{hagedorn}:
\begin{equation}
\frac{1}{N_{ev}} \cdot \frac{d^{2}N}{dp^{2}_{T} d\eta}=
exp(a - b \cdot \sqrt{p_{T}^{2} + m_{\pi}^{2}})
\label{e:exp}
\end{equation}
where $m_{\pi}$ is the pion mass.
The results of the fits of this function to ZEUS data in the interval
$0.3<p_{T}<1.2\GeV$ are also shown as the full line
in Fig.~\ref{f:corrected_pt}.
The resulting values of the exponential slope $b$ are listed in
Table \ref{t:slopes}.
The systematic errors were estimated by varying the
relative inclusive cross sections within the systematic error limits
(see section \ref{s:systematics}) and by varying the upper
boundary of the fitted interval from $p_{T}=1.0\GeV$ to $1.4\GeV$.

\begin{figure}[h]
\centerline{\hbox{
\psfig{figure=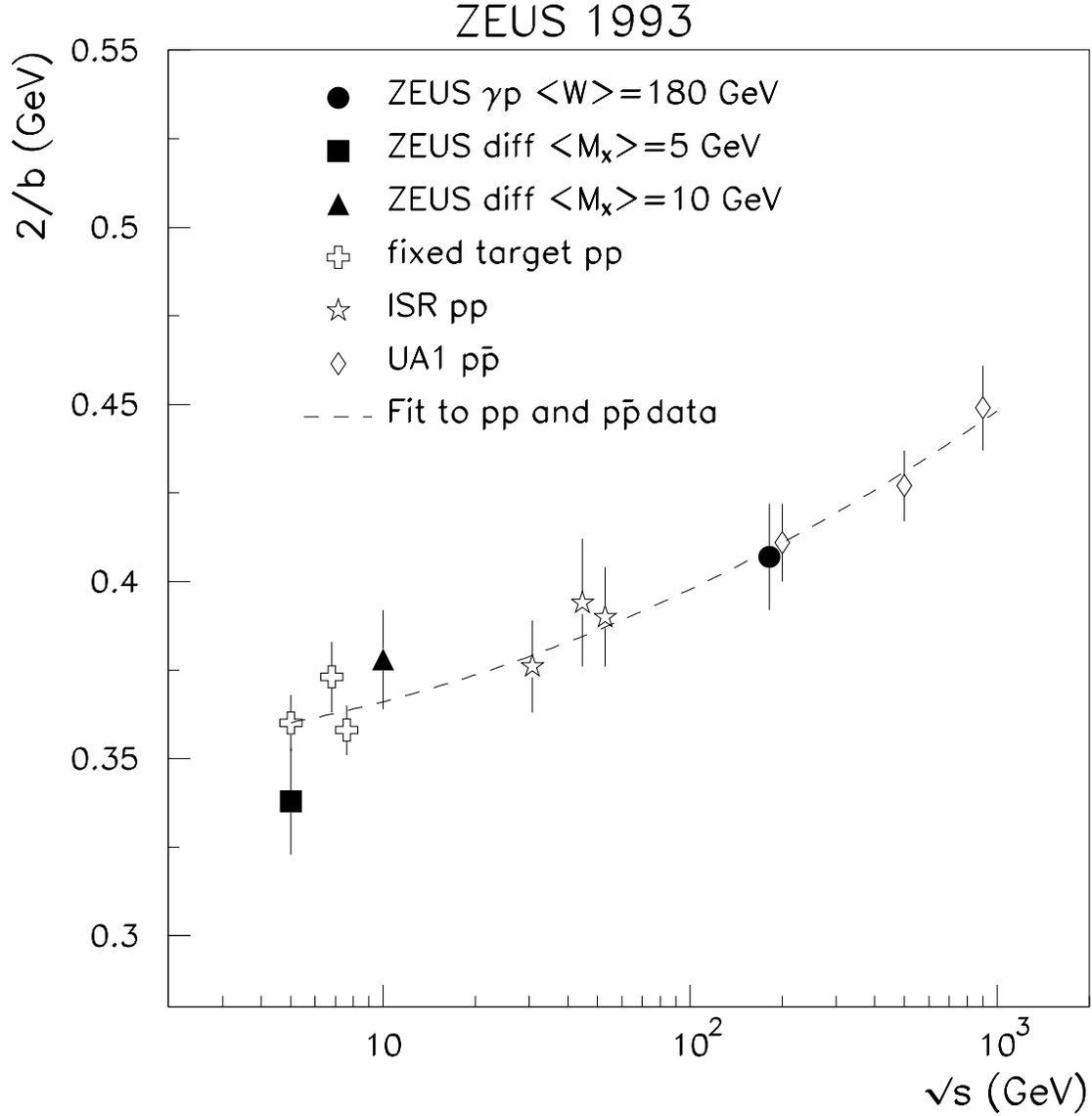}}}
\bf\caption{\it
Inverse slope of the
exponential fit of the form (\protect\ref{e:exp})
vs. the c.m. energy for different experiments.
The fixed target data were taken from \protect\cite{72A1}
\protect\cite{67A1} \protect\cite{71A1}, ISR -- \protect\cite{ISR},
UA1-- \protect\cite{UA1}.
The ZEUS non--diffractive point is placed at the
photon--proton c.m. energy.
The diffractive points are plotted at the energies corresponding
to the mean value of the invariant mass $\langle M_{X} \rangle$.
The error bars indicate the quadratic sum of
the statistical and systematic errors.
The dashed line is a parabola in $log(s)$ and was fitted to
all the hadron--hadron points to indicate the trend of the data.
}
\label{f:pt_slopes}
\end{figure}

In Fig.~\ref{f:pt_slopes} we present a comparison of the
$b$ parameter resulting from the fits of (\ref{e:exp}) to proton-proton and
proton-antiproton data as a function of the c.m. energy.  The
slope of the ZEUS non-diffractive spectrum agrees
with the data from hadron--hadron scattering at an energy close to the ZEUS
photon--proton c.m. energy.
The diffractive slopes agree better with the hadronic data corresponding
to a lower energy.  In Fig.~\ref{f:pt_slopes}
the ZEUS diffractive points are plotted at $5\GeV$ and $10\GeV$,
the values of the invariant mass of the dissociated photon.
A similar behaviour has been observed
for the diffractive dissociation of protons, i.e.
the scale of the fragmentation of the excited system is
related to the invariant mass rather than to the total c.m.
energy \cite{p-diff}.
The dashed line in Fig.~\ref{f:pt_slopes} is a parabola in $log(s)$
and was fitted to all the hadron--hadron points to indicate
the trend of the data.  As one can see, our photoproduction results
are consistent with the hadronic data.

\begin{table}[h]
\centerline{\hbox{
\begin{tabular}{|c||c|c|c|c|c|c|}
\hline
\ \ sample \ \ &
 $b [\GeV^{-1}]$ &
 $\sigma_{stat}(b)$ &
 $\sigma_{syst}(b)$ &
 $a$ &
 $\sigma_{stat}(a)$ &
 $cov(a,b)$\\
\hline\hline
non-diffractive &
4.94 & 0.09 & 0.19 & 3.39 & 0.09 & -0.011\\
diff $\langle M_{X} \rangle = 5\GeV$ &
5.91 & 0.17 & 0.19 & 2.78 & 0.10 & -0.016\\
diff $\langle M_{X} \rangle = 10\GeV$ &
5.28 & 0.10 & 0.17 & 3.34 & 0.06 & -0.006\\
\hline
\end{tabular}
}}

\vspace{1cm}
\bf\caption{\it
The values of the parameters resulting from the fits of equation
(\protect\ref{e:exp}) to ZEUS data in the interval $0.3<p_{T}<1.2\GeV$.
The $\sigma_{stat}$ and $\sigma_{syst}$ indicate the statistical and
systematic errors.}
\label{t:slopes}
\end{table}

The non--diffractive spectrum in Fig.~\ref{f:corrected_pt}
clearly departs from the
exponential shape at high $p_{T}$ values.  Such a behaviour is expected
from the contribution of the hard scattering of partonic constituents
of the colliding particles, a process that can be described in the framework
of perturbative QCD.  It results in a high $p_{T}$ behaviour of the
inclusive spectrum that can be approximated by a
power law formula:
\begin{equation}
\frac{1}{N_{ev}} \cdot \frac{d^{2}N}{dp^{2}_{T} d\eta}=
  A \cdot (1 + \frac{p_{T}}{p_{T\:0}})^{-n}
\label{e:power}
\end{equation}
where $A$, $p_{T\:0}$ and $n$ are parameters determined from the data.
The fit to the ZEUS points in the region of $p_{T}>1.2\GeV$
gives a good description of the data and
results in the parameter values
 $p_{T\:0}=0.54$ GeV, $n=7.25$ and $A=394$ GeV$^{-2}$.
The statistical precision of these numbers is
described by the covariance matrix shown in Table \ref{t:cov}.
The fitted function
is shown in Fig.~\ref{f:corrected_pt} as the dotted line.

\begin{table}[h]
\centerline{\hbox{
\begin{tabular}{|c||c|c|c|}
\hline
  & $p_{T\:0}$ & $n$ & $A$ \\
\hline\hline
 $p_{T\:0}$ & $0.32\cdot 10^{-3}$ & $0.48\cdot 10^{-3}$ & $-0.10\cdot 10^{1}$\\
        $n$ &                     & $0.12\cdot 10^{-2}$ & $-0.12\cdot 10^{1}$\\
   $A$      &                    &                      & $ 0.32\cdot 10^{4}$\\
\hline
\end{tabular}}}
\vspace{1cm}
\bf\caption{\it
The covariance matrix corresponding to the fit of
equation (\protect\ref{e:power}) to the non--diffractive data
for $p_{T}>1.2\GeV$.}
\label{t:cov}
\end{table}

\begin{figure}[p]
\centerline{\hbox{
\psfig{figure=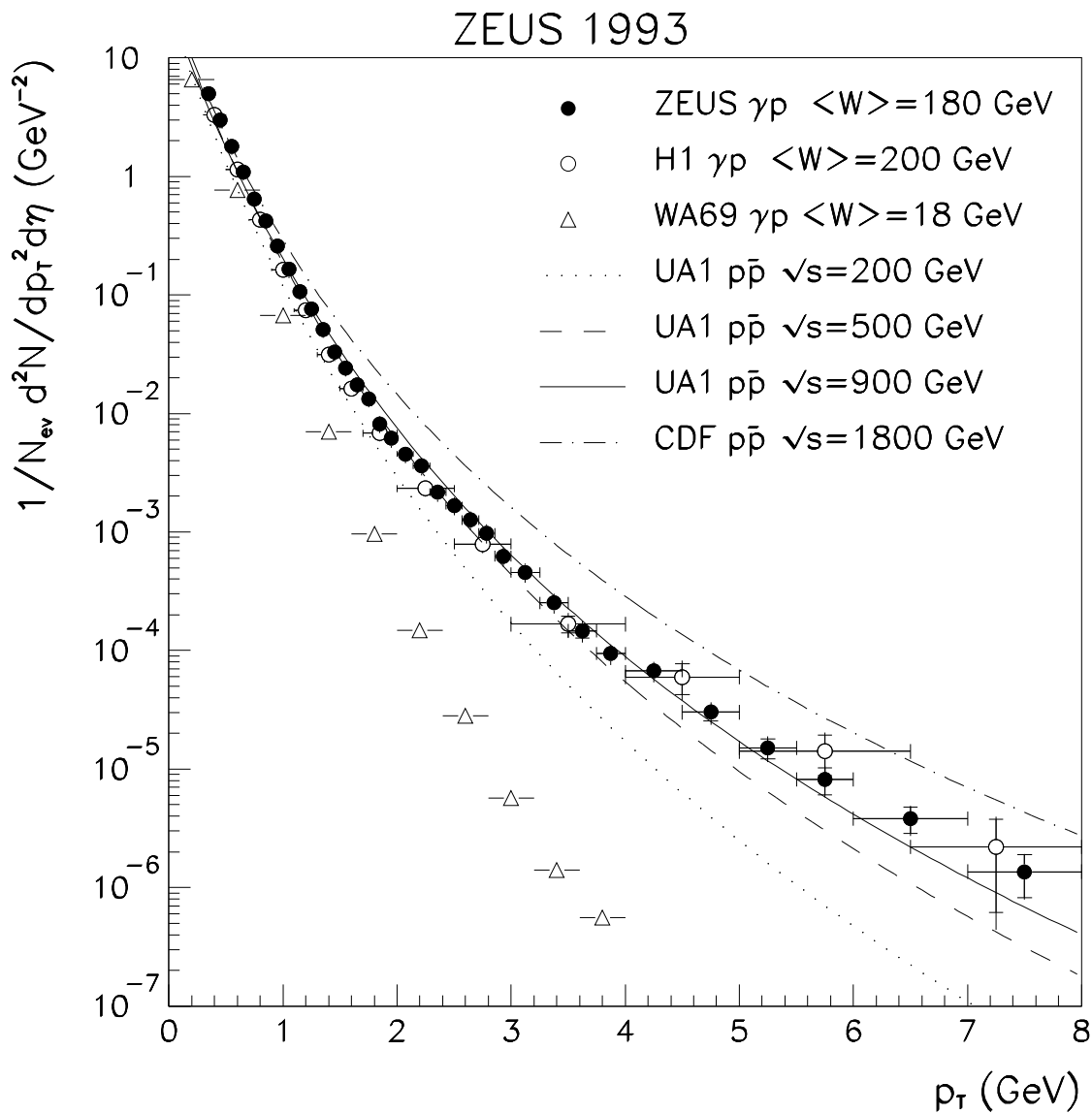}}}
\bf\caption{\it
Comparison of ZEUS non--diffractive transverse momentum spectrum with
the data from
H1 \protect\cite{H1},
OMEGA \protect\cite{omega},
UA1 \protect\cite{UA1} and
CDF \protect\cite{CDF}.
The inner error bars indicate the
statistical errors and the outer ones represent
the quadratic sum of the statistical and
systematic errors.  }
\label{f:pt_comparison}
\end{figure}

In Fig.~\ref{f:pt_comparison} the ZEUS data are presented together
with the results of a similar measurement from the H1 collaboration at
$\langle W \rangle = 200$ GeV \cite{H1} and
the data from the WA69 photoproduction experiment at a
c.m. energy of $\langle W \rangle = 18\GeV$ \cite{omega}.
For the purpose of this comparison, the inclusive cross sections
published by those experiments
were divided by the corresponding total photoproduction cross sections
\cite{H1-sigmatot,ALLM}.
Our results are in agreement with the H1 data.
The comparison with the WA69 data shows that
the transverse momentum spectrum becomes harder as the
energy of the $\gamma p$ collision increases.
Figure~\ref{f:pt_comparison} also shows the functional
fits of the form (\ref{e:power}) to {\it p\={p}} data from UA1 and CDF
at various c.m. energies \cite{UA1,CDF}.
Since the fits correspond to inclusive cross sections published by
these experiments, they have been divided by the
cross section values used by these experiments for the absolute
normalization of their data.
The inclusive $p_{T}$ distribution from our
photoproduction data is clearly harder than the distribution
for {\it p\={p}} interactions at a similar c.m. energy
and in fact is similar to {\it p\={p}} at $\sqrt{s}=900\GeV$.

This comparison indicates that in spite of the apparent similarity
in the low $p_{T}$ region
between photoproduction and proton--antiproton collisions at a similar
c.m. energy, the two reactions are different in the hard regime.
There are many possible reasons for this behaviour.
Firstly, both of the {\it p\={p}} experiments used for the comparison
measured the central rapidity region
($|\eta|<2.5$ for UA1 and $|\eta|<1$ for CDF),
while our data correspond to $0.8<\eta_{c.m.}<3.4$.
Secondly,
according to VDM, the bulk of the $\gamma p$ collisions can be approximated as
an interaction of a vector meson $V$ with the proton.  The $p_{T}$ spectrum
of $Vp$ collisions may be harder than {\it p\={p}}
at a similar c.m. energy, since the parton momenta of quarks
in mesons are on average larger than in baryons.
Thirdly, in the picture where the photon consists of a resolved part
and a direct part, both the anomalous component of the resolved
photon and the direct photon become significant at high
$p_{T}$ and make the observed spectrum harder compared to that
of $Vp$ reactions.

\begin{figure}[h]
\centerline{\hbox{
\psfig{figure=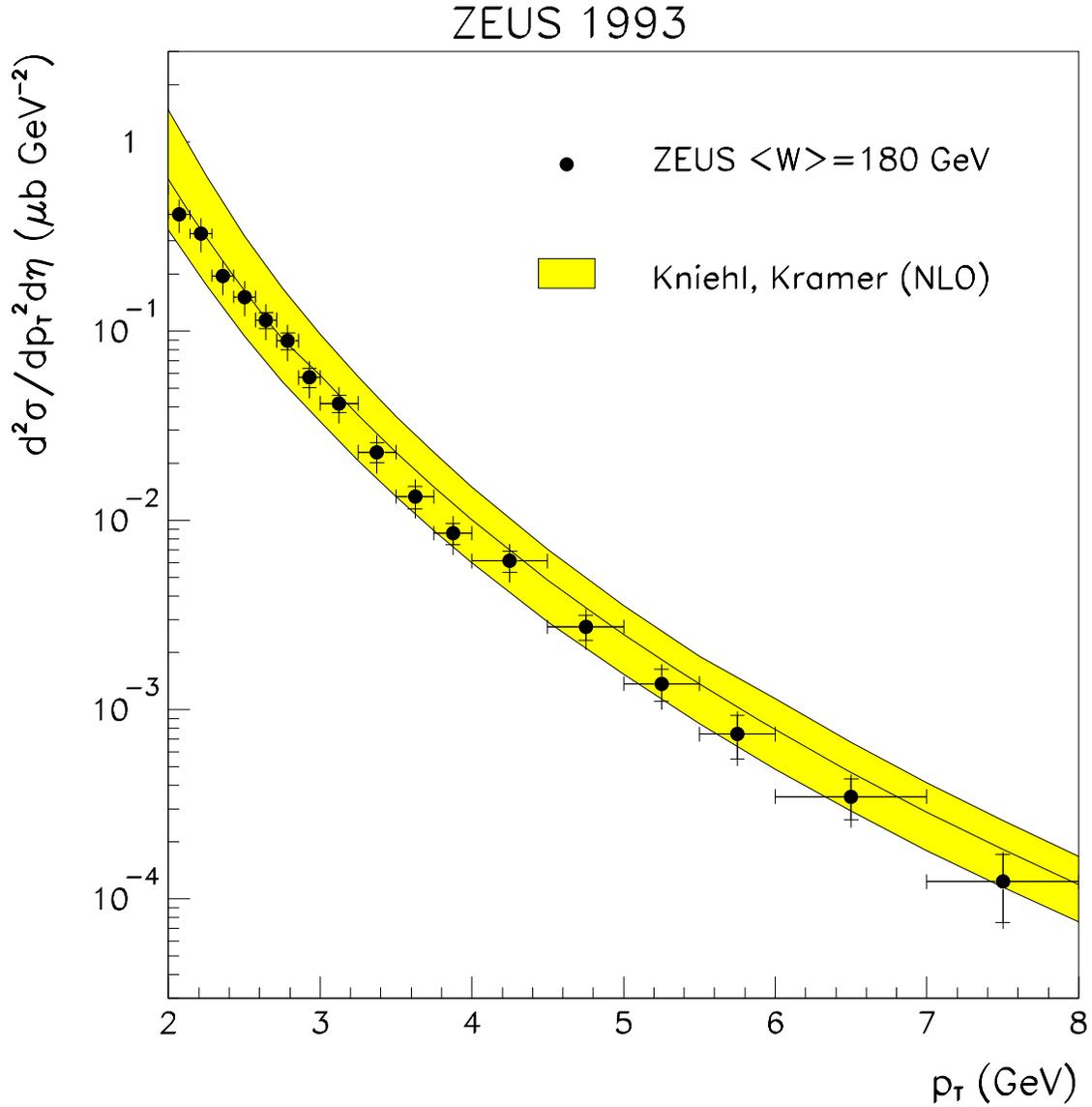}}}
\bf\caption{\it Comparison of the
ZEUS inclusive cross sections for non--diffractive
photoproduction
at $\langle W_{\gamma p}\rangle = 180\GeV$ with the NLO QCD calculation
results from \protect\cite{krammer}.
The inner error bars on the data points indicate
the statistical errors and the outer ones represent
the quadratic sum of the statistical and
systematic errors.  The shaded band corresponds to the uncertainty of
the theoretical calculation.
}
\label{f:pt_kniehl}
\end{figure}

Figure \ref{f:pt_kniehl} shows the comparison of our non--diffractive data
with the theoretical prediction obtained recently
from NLO QCD calculations \cite{krammer}.
The charged particle production rates in a non--diffractive event
were converted to inclusive non-diffractive cross sections
by multiplying by the non--diffractive
photoproduction cross section of
$\sigma_{nd}(\gamma p \rightarrow X)=91\pm 11{\rm \:\mu b}$
\cite{zeus-sigmatot}.
The theoretical calculations relied on
the GRV parametrisation of the parton densities in the photon
and on the CTEQ2M parametrisation for partons in the proton\cite{CTEQ}.
The NLO fragmentation functions describing the relation between the
hadronic final state and the partonic level
were derived from the $e^{+}e^{-}$ data \cite{krammer-fragm}.
The calculation depends strongly on the parton densities in the
proton and in the photon, yielding a spread in the
predictions of up to $30\%$ due to the former and $20\%$ due to the latter.
The factorization scales of the incoming and outgoing parton lines,
as well as the renormalization scale,
were set to $p_{T}$.  The uncertainty due to the
ambiguity of this choice was estimated by changing all three
scales up and down by a factor of 2.
The estimates of the theoretical errors were added in quadrature and
indicated in Fig.~\ref{f:pt_kniehl} as a shaded band.
The theoretical calculation is in good agreement with the
ZEUS data.

\section{Conclusions}
We have measured the inclusive transverse momentum spectra of charged particles
in diffractive and non--diffractive
photoproduction events with the ZEUS detector.
The inclusive transverse momentum spectra fall
exponentially in the low $p_{T}$ region, with a slope that
increases slightly going
from the non--diffractive to the diffractive collisions with the lowest
$M_{X}$.
The diffractive slopes are consistent with hadronic data at a c.m. energy
equal to the invariant mass of the diffractive system.
The non--diffractive low $p_{T}$ slope is consistent with the
result from {\it p\={p}} at a similar c.m. energy but
displays a high $p_{T}$ tail
clearly departing from the exponential shape.
Compared to photoproduction data at a lower c.m. energy we observe
a hardening of the transverse momentum spectrum as the collision energy
increases.
The shape of our $p_{T}$ distribution is comparable to that of
{\it p\={p}} interactions at  $\sqrt{s}=900\GeV$.
The results from a NLO QCD calculation agree with the measured
cross sections for inclusive charged particle production.

\section{Acknowledgments}

We thank the DESY Directorate for their strong support and
encouragement.  The remarkable achievements of the HERA machine group were
essential for the successful completion of this work and are gratefully
appreciated.  We gratefully acknowledge the support of the DESY computing
and network services.
We would like to thank B.A. Kniehl and G. Kramer for useful discussions
and for providing the NLO QCD calculation results.

\pagebreak

\end{document}